\documentclass[summary]{ursi}

\usepackage{subfigure}

\title{Experimental Characterization of ISAC Channel Mapping and Environment Awareness}

\author{Zhuangzhuang Cui*\affref{ref1}, Rizqi Hersyandika\affref{ref2}, Haoqiu Xiong\affref{ref1},
  and Sofie Pollin\affref{ref1}\affref{ref3}}

\affiliation{%
  \aff{ref1}{WaveCoRE, Department of Electrical Engineering (ESAT), KU Leuven, 3000 Leuven, Belgium}
  \aff{ref2}{Department of Electrical Engineering, Eindhoven University of Technology (TU/e), 5600 MB Eindhoven,
Netherlands}
\aff{ref3}{Interuniversity Microelectronics Centre (imec), 3001 Leuven, Belgium.}
}

\begin{document}

\maketitle

\begin{abstract}
In the context of integrated sensing and communications (ISAC), this paper presents an experimental investigation of the relationship between monostatic sensing and naturally bistatic communication channels in an indoor millimeter-wave environment. We characterize the propagation channel in the joint delay--angle domain, extract dominant multipath components (MPCs) and associate them with physical scatterers in the environment, and demonstrate how communication MPCs can be explicitly recovered from sensing channels. Finally, the radar cross-sections (RCSs) of two key scatterers, namely the wall and metal plate, are obtained based on calibrated channel power and reconstructed propagation distances.
\end{abstract}

\section{Introduction}

Integrated sensing and communications (ISAC) has emerged as a key enabling technology for sixth-generation (6G) wireless systems. The fundamental vision of ISAC is to unify radio sensing and data transmission on a shared hardware and waveform platform, thereby enabling joint perception and connectivity through the same radio infrastructure \cite{isacoverview}. In this paradigm, wireless channels are no longer treated as black-box stochastic processes, but rather as physical propagation phenomena governed by geometry and electromagnetic interactions with the surrounding environment.

From a topology perspective, one of the most fundamental ISAC configurations is monostatic sensing, where sensing functionality is performed at the communication transceiver itself. In such a setup, the transceiver actively probes the environment and receives echoes from surrounding objects, while simultaneously supporting data transmission. This naturally leads to a close coupling between sensing and communication channels, since both are shaped by the same physical scatterers. Nevertheless, the two channels generally exhibit different visibility of the environment due to differences in the field of view, antenna beam patterns, and monostatic versus bistatic propagation geometries.

Recent work has demonstrated that when communication scatterers are part of sensing targets, the communication channel can be explicitly recovered from monostatic sensing measurements \cite{cuiisac}. Dense angular scanning enables the extraction of multipath components (MPCs) and their association with physical reflectors, supporting environment-aware channel modeling. In parallel, 3GPP standardization activities have emphasized the importance of target and background channels in bistatic sensing and communication scenarios, where radar cross-section (RCS) plays a central role in characterizing sensing reflections \cite{3GPP}. Geometry-based stochastic channel models further explore statistical correlations between sensing and communication links through shared clusters and scatterers \cite{liubupt}, while recent terahertz monostatic measurements demonstrated geometry-resolved propagation modeling and the corresponding scene reconstruction \cite{thz}.

Motivated by these developments, this paper presents an experimental study combining dense monostatic angular scanning and bistatic communication channel measurements. By explicitly identifying the propagation geometry, we establish a deterministic linkage between sensing and communication channels and demonstrate environment-aware ISAC channels with physical interpretability.

\section{Setup and Experiments}

\subsection{Measurement Setup}

The measurement campaign is conducted using a standard millimeter-wave channel sounding system operating at 62.64~GHz with a bandwidth of 1.76~GHz. The system follows the IEEE 802.11ay protocol, where channel estimation is performed using complementary Golay sequences of length 128 embedded in the preamble. This results in a frequency resolution of 13.75~MHz. The zero-sidelobe autocorrelation property of Golay sequences enables an accurate estimation of the channel impulse response (CIR) \cite{channelest_80211}.

\begin{figure}[t]
  \centering
  \subfigure[]{
  \includegraphics[width=60mm]{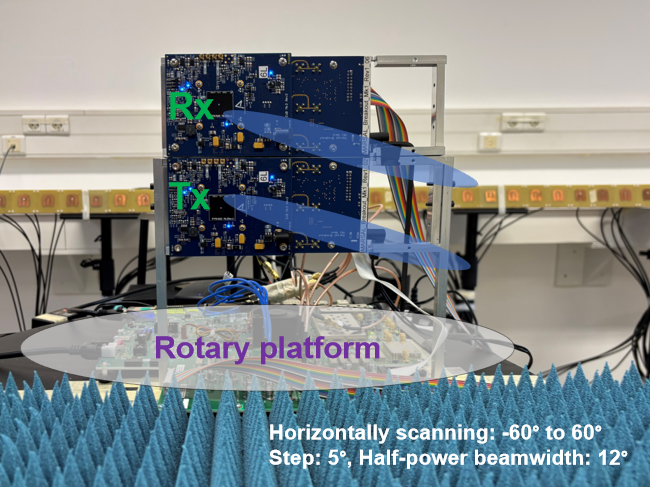}}
  \subfigure[]{\includegraphics[width=60mm]{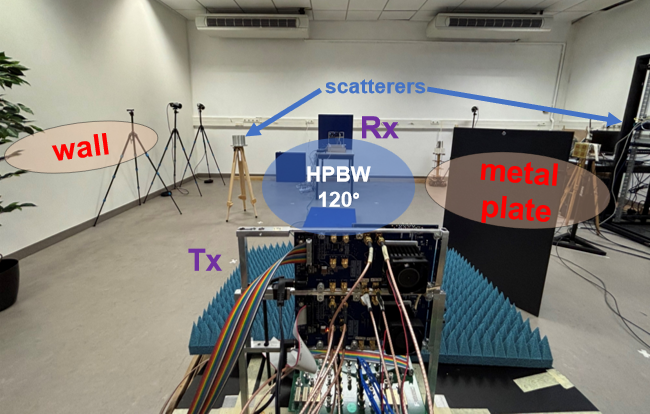}}
  \caption{Indoor ISAC measurement setup revised from \cite{cuiisac}: (a) monostatic scanning; (b) bistatic communication.}
  \label{fig:setup}
\end{figure}

Monostatic and bistatic configurations are deployed in an indoor environment. In the monostatic setup, the transmitter (Tx) and receiver (Rx) are co-located on a rotating platform equipped with radio-frequency absorbers to suppress platform reflections, as shown in Fig.~\ref{fig:setup}(a). The platform enables precise angular scanning over $[-60^\circ,60^\circ]$ with a step size of $5^\circ$. An $8\times8$ antenna array with a half-power beamwidth (HPBW) of $12^\circ$, steered in the boresight direction, is used to achieve a high angular resolution. 



In the bistatic configuration, Tx and Rx are separated in a LoS distance of $d_{\rm LoS}=4$~m and placed at the two aforementioned monostatic locations, as shown in Fig.~\ref{fig:setup}(b). The Rx employs a wide beam with $120^\circ$ HPBW to capture a quasi-omnidirectional communication channel. The environment contains several dominant reflectors, including a wall, a metal plate, a cylindrical reflector, and a few metallic frames, which contribute to the observed MPCs.

\subsection{Power--Angle--Delay Profiles}

\begin{figure}[tbp]
  \centering
  \subfigure[]{
  \includegraphics[width=60mm]{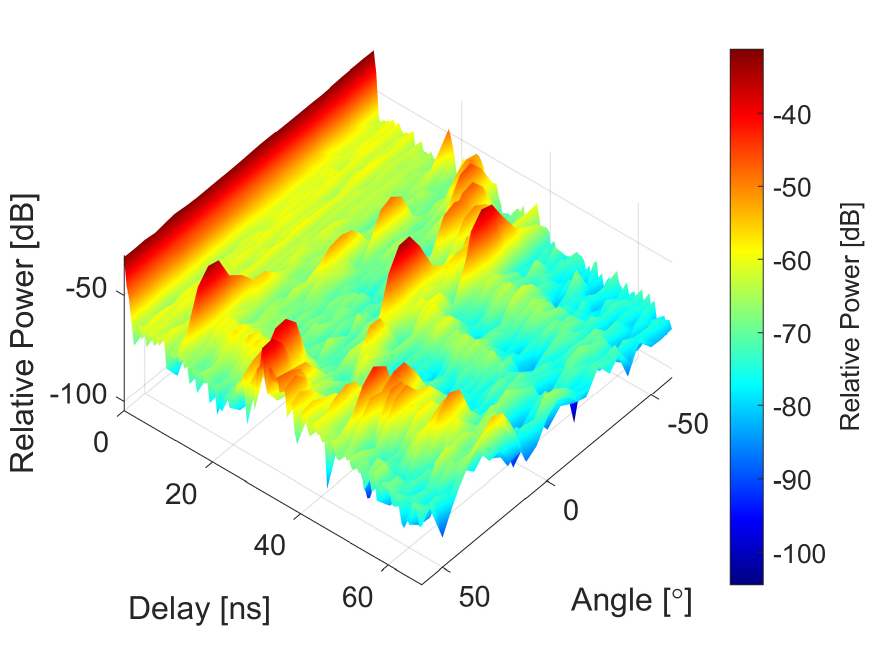}}
  \subfigure[]{
  \includegraphics[width=60mm]{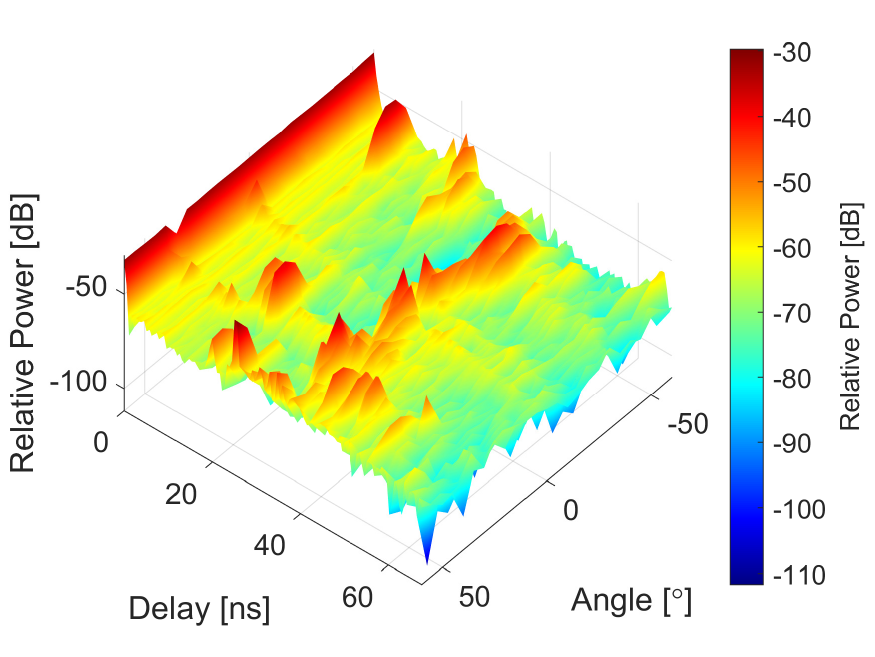}}
  \caption{Calibrated PADPs measured at (a) monostatic location $\text{Mono}1$ and (b) monostatic location $\text{Mono}2$.}
  \label{fig:padp}
\end{figure}

The propagation characteristics of the environment are first investigated through monostatic angular scanning. Two monostatic setup placements, referred to as $\text{Mono}_1$ and $\text{Mono}_2$, are considered to capture the channel from opposite directions. Specifically, the $\text{Mono}_1$ and $\text{Mono}_2$ setups are positioned at the Tx and Rx locations, respectively, of the bistatic configuration shown in Fig.~\ref{fig:setup}(b). Fig.~\ref{fig:padp} shows the calibrated three-dimensional power--angle--delay profiles (PADPs) measured at the two monostatic locations. The delay is normalized with respect to the strongest path.

A dominant direct-coupling component is observed at zero excess delay, corresponding to the leakage path between the co-located transmitter and receiver. This component is followed by several strong MPCs with excess delays extending up to approximately 75~ns. These MPCs exhibit clear angular separations, indicating that the multipath structure is dominated by a small number of scatterers in the environment. The narrow beamwidth of $12^\circ$ enables accurate angular discrimination of closely spaced scatterers, resulting in well-resolved MPCs in the joint angle--delay domain.

\begin{figure}[tbp]
  \centering
  \includegraphics[width=60mm]{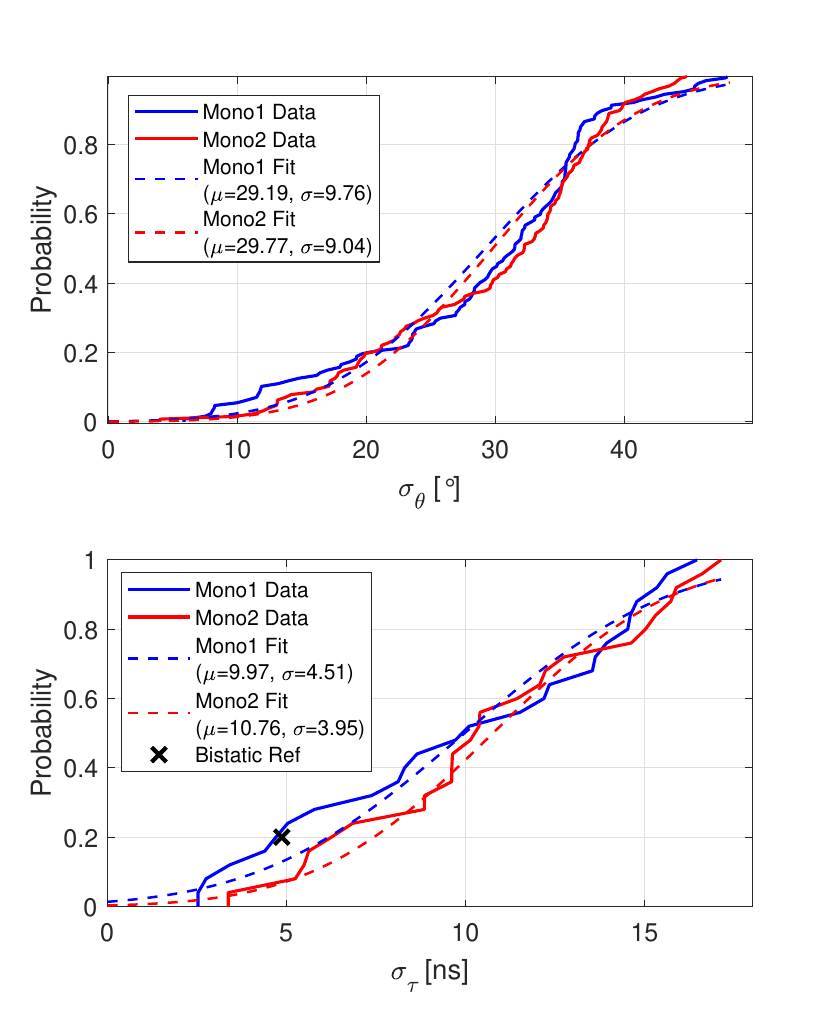}
  \caption{CDFs of RMS angular/delay spreads and their Normal fitted results for two monostatic locations.}
  \label{fig:CDF}
\end{figure}

To further quantify the multipath dispersion, the root-mean-square (RMS) delay spread $\sigma_\tau$ is computed per steering angle, whereas the RMS angular spread $\sigma_\theta$ is computed per delay bin from the PADP. Fig.~\ref{fig:CDF} shows the cumulative distribution functions (CDFs) for both monostatic locations.

The angular spread exhibits a mean value of approximately $29^\circ$ with standard deviations of $9.76^\circ$ and $9.04^\circ$ for $\text{Mono}1$ and $\text{Mono}2$, respectively, indicating a moderately rich angular scattering environment despite the dominance of specular reflections. The delay spread is concentrated in the range of 5--15~ns, with mean values of 9.97~ns ($\text{Mono}1$) and 10.76~ns ($\text{Mono}2$). A reference point corresponding to the bistatic communication link is also shown, with $\sigma_\tau$ of 4.86~ns, which lies within the monostatic distributions but is smaller due to the lower antenna gain of the wide-beam configuration, which limits the detection of weaker paths.

\section{ISAC Channels }
\subsection{Extraction of Monostatic Scatterers}
Let $h(\theta,\tau)$ denote the calibrated monostatic CIR measured at steering angle $\theta \in \{\theta_a\}$, where $\theta_a$ spans $[-60^\circ,60^\circ]$ with $5^\circ$ steps. For each angular snapshot $h(\theta_a,\tau)$, local maxima are detected with a minimum peak height threshold $P_{\min}$ (-55 dB) and a minimum peak separation $\Delta\tau_{\min}$ (2.2~ns) to avoid multiple detections within the same resolvable delay bin. To prevent spurious detections from near-field leakage or residual coupling around zero delay, a range gate is applied such that only peaks with range $R \geq R_{\min}$ (0.5~m) are retained. In addition, at most $K_{\max}$ (set as 4) peaks are kept per angle to focus on the dominant MPCs.

\begin{figure}[tbp]
  \centering
  \includegraphics[width=60mm]{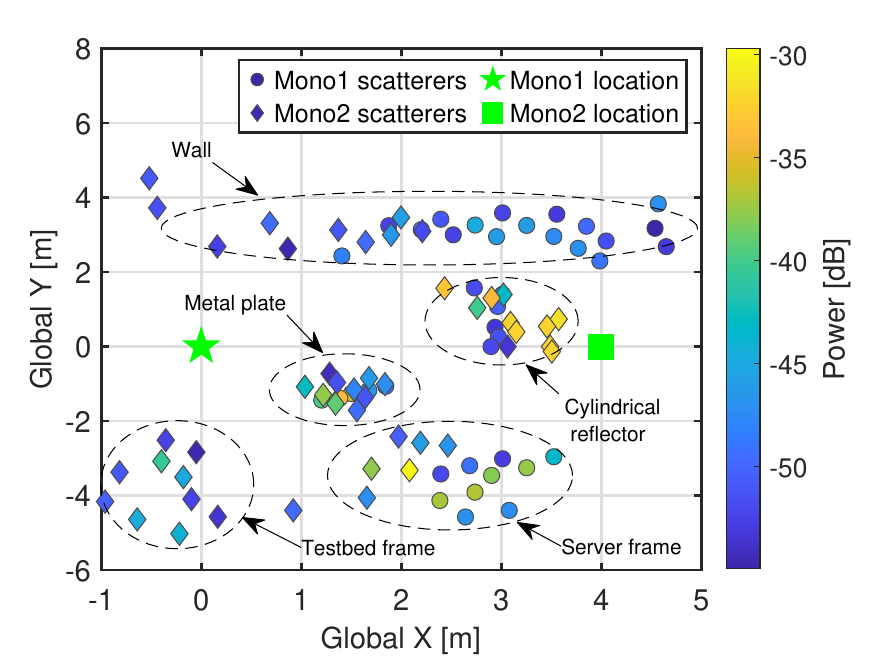}
  \caption{Extracted MPCs from two monostatic locations.}
  \label{fig:scatterer_map}
\end{figure}

Each detected peak at delay index $\ell$ is converted to an equivalent one-way path length (range) $R(\ell)$ using the calibrated delay-to-range mapping,
$R(\ell)=\frac{c}{2}\tau(\ell)$,
where $c$ is the speed of light and the factor $1/2$ accounts for the monostatic round trip. The corresponding scatterer location is then obtained by geometric back-projection in the global coordinate system. For $\text{Mono}_1$ located at $p_1=[x_1,y_1]^T$, the scatterer point associated with $(\theta_a,R)$ is
\begin{equation}
\mathbf{m}_{1}(\theta_a,R)=
\begin{bmatrix}
x_1 + R\cos\theta_a\\[2pt]
y_1 - R\sin\theta_a
\end{bmatrix},
\label{eq:bp_mono1}
\end{equation}
and for $\text{Mono}_2$ located at $p_2=[x_2,y_2]^T$,
\begin{equation}
\mathbf{m}_{2}(\theta_a,R)=
\begin{bmatrix}
x_2 - R\cos\theta_a\\[2pt]
y_2 - R\sin\theta_a
\end{bmatrix},
\label{eq:bp_mono2}
\end{equation}
where the sign difference in the $x$-coordinate reflects the opposite-facing direction of the two monostatic locations.
Fig.~\ref{fig:scatterer_map} shows the reconstructed scatterer map in the global coordinate system, where the $\text{Mono}1$ location is set as the origin $(0,0)$ and $\text{Mono}2$ is located at $(4,0)$. Five dominant scattering clusters are clearly identified: 1) wall, 2) metal plate, 3) cylindrical reflector, 4) server frame, and 5) testbed frame, confirming the main objects in the environment, partly seen in Fig.~\ref{fig:setup}(b). As horizontal-plane scanning is employed, we capture only scatterers exposed near the transceiver height ($\approx1$~m). 

\subsection{Association of Communication MPCs}
\begin{figure}[tbp]
  \centering
  \includegraphics[width=60mm]{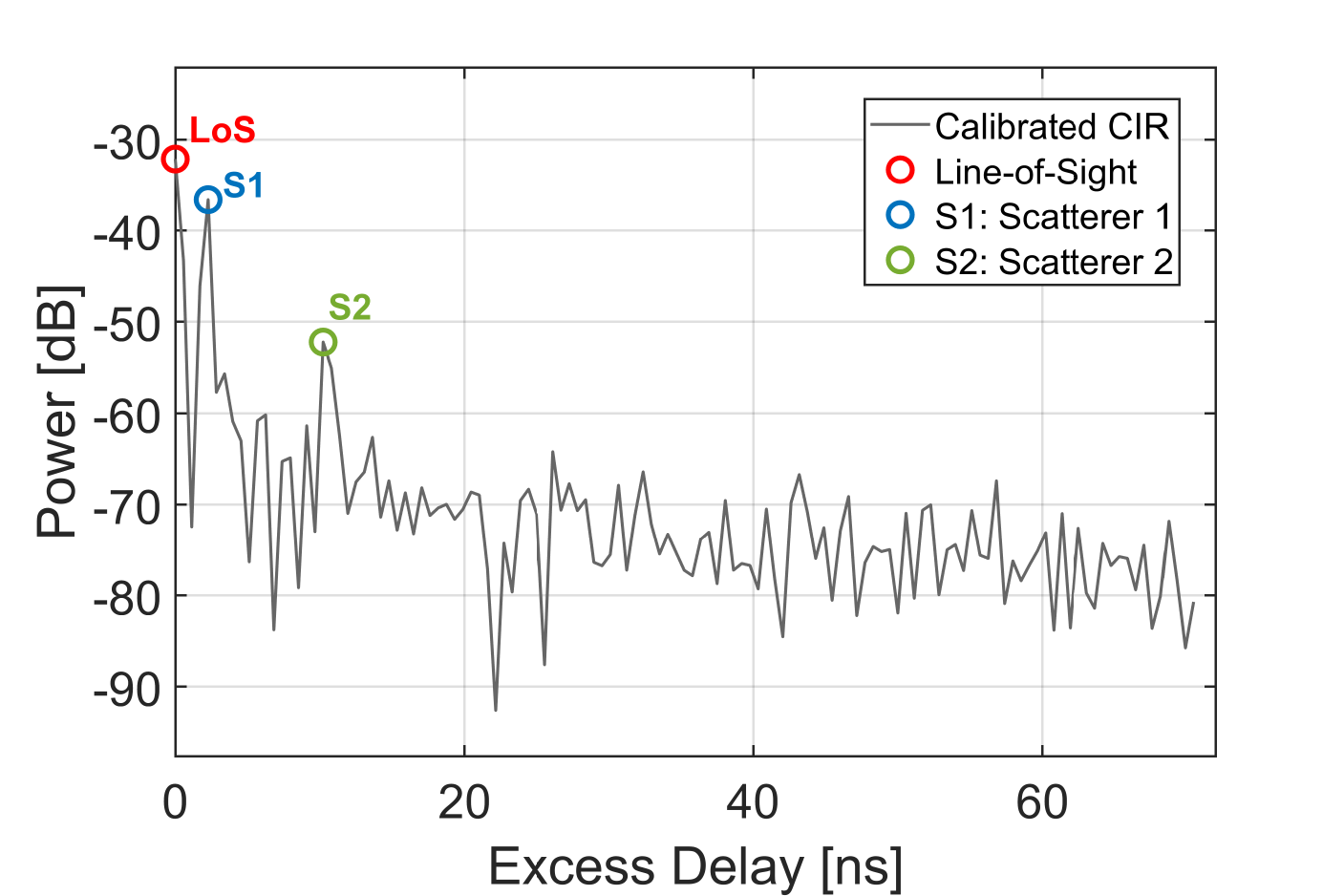}
  \caption{Communication channel with components LoS, S1 and S2 under normalized delay.}
  \label{fig:bichannel}
\end{figure}
Fig.~\ref{fig:bichannel} shows the calibrated bistatic CIR. In addition to the LoS component, two dominant MPCs, denoted as S1 and S2, are observed at excess delays of approximately 2.27~ns and 10.22~ns, respectively.
The physical origin of S1 and S2 is determined by comparing their measured excess delays with the path lengths implied by the sensing-derived scatterer map in Fig.~\ref{fig:scatterer_map}. For a candidate scatterer point $\mathbf{m}$, the predicted bistatic excess delay is
\begin{equation}
\Delta \tau(\mathbf{m}) = \frac{\|\mathbf{m}-p_t\|+\|\mathbf{m}-p_r\|-d_{\rm LoS}}{c},
\label{eq:bistatic_delay_pred}
\end{equation}
where $p_t$ and $p_r$ denote the bistatic locations at $p_1$ and $p_2$, respectively. Evaluating \eqref{eq:bistatic_delay_pred} over the dominant sensing clusters shows that the delay of S1 matches the metal plate, while S2 matches the wall, establishing that the dominant bistatic MPCs are physically contained within the monostatic channels, showing a deterministic geometric correspondence between sensing and communication channels.

\begin{figure}[tbp]
  \centering
  \subfigure[]{
  \includegraphics[width=60mm]{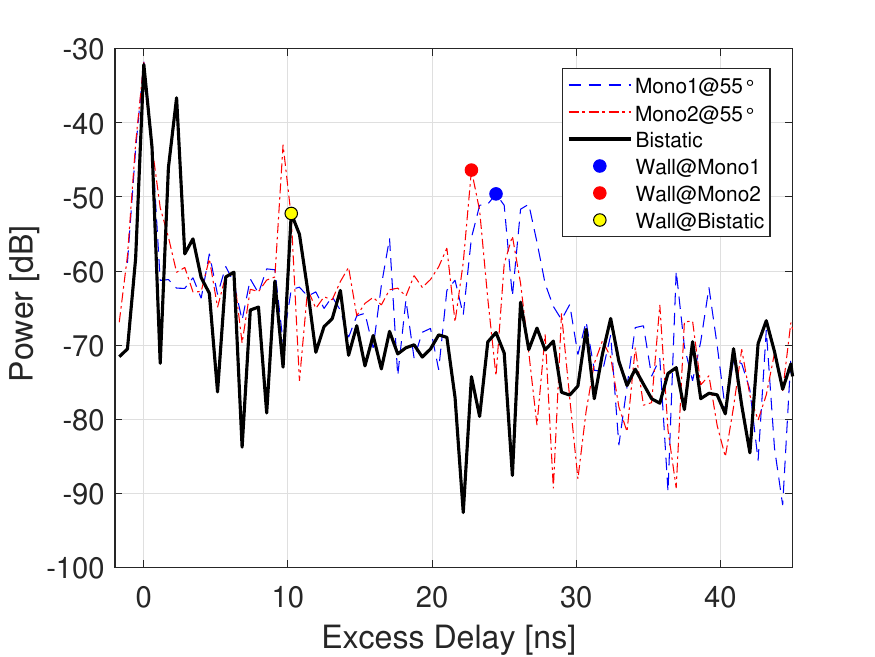}}
  \subfigure[]{
  \includegraphics[width=65mm]{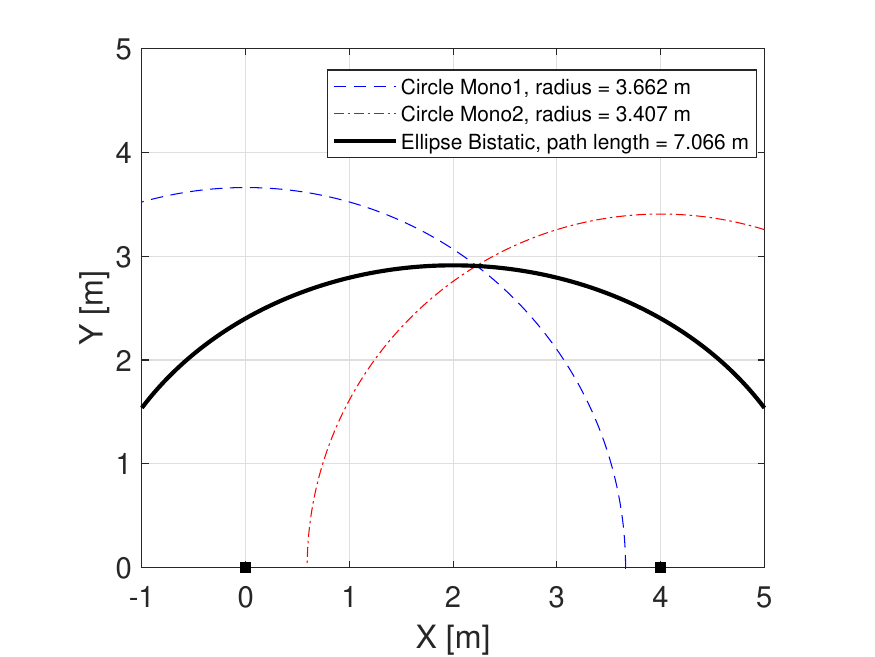}}
  \caption{Wall reflection in sensing and communication channels: (a) detection and (b) geometric identification.}
  \label{fig:wall}
\end{figure}

\subsection{ Key Scatterers}
Building on the scatterer mapping and sensing--communication association, we now focus on the physical identification and characterization of the dominant scatterers.
Fig.~\ref{fig:wall} illustrates the detection and geometric reconstruction of the wall reflection. The reconstructed monostatic sensing circles with radii of 3.662~m ($\text{Mono}1$) and 3.407~m ($\text{Mono}2$) at $\theta_a=55^\circ$, and a bistatic ellipse with a total path length of 7.066~m, intersect the physical wall with a localization error of 3~mm, providing strong validation of the physical origin of the detected reflection.

With the propagation geometry validated, we proceed to estimate the RCSs of the dominant scatterers using a LoS-referenced calibration strategy based on the bistatic communication link. The bistatic LoS component provides a power reference via the Friis free-space relationship, while the monostatic sensing channel, affected by direct coupling, is used primarily for accurate range reconstruction.
Starting from the bistatic radar equation $P_{\rm ref}= P_t G_t G_r \frac{\lambda^2 \sigma_b}{(4\pi)^3 R_t^2 R_r^2}$
and the Friis free-space relation for the bistatic LoS reference $P_{\rm LoS}= P_t G_t G_r \left(\frac{\lambda}{4\pi d_{\rm LoS}}\right)^2$, we then can take the ratio to eliminate the unknown constant terms $(P_t,G_t,G_r)$. The bistatic path length is recovered from the excess delay relative to the LoS reference as $D_{\rm bi}=d_{\rm LoS}+c(\tau_{\rm ref}-\tau_{\rm LoS})$, while the transmitter-to-target distance is obtained from the monostatic excess delay as $R_t=\frac{c\tau_{\rm mono}}{2}$, and $R_r=D_{\rm bi}-R_t$. The bistatic RCS is then extracted as
\begin{equation}
\begin{aligned}
\sigma_{b,\mathrm{dBsm}} &=
\big(P_{\rm ref}-P_{\rm LoS}\big)
+ 20\log_{10}(R_t)+20\log_{10}(R_r) \\
&\quad - 20\log_{10}(d_{\rm LoS})
+ 10\log_{10}(4\pi),
\end{aligned}
\end{equation}
yielding $\sigma_{\rm wall}=0.76$~dBsm and $\sigma_{\rm plate}=8.73$~dBsm. The significantly larger RCS of the metal plate is attributed to its smooth conductive surface and strong specular reflection, whereas the wall exhibits a lower RCS due to its rough dielectric surface and increased diffuse scattering.

\section{Conclusion}
This paper presented an experimental characterization of monostatic sensing and bistatic communication channels in an indoor ISAC scenario. Dense monostatic scanning reveals a sparse propagation structure dominated by a small number of scatterers whose locations can be explicitly reconstructed. The dominant MPCs of bistatic communication are shown to be physically contained within the sensing channels. Furthermore, the RCSs of the wall and the metal plate were experimentally measured as $0.76$~dBsm and $8.73$~dBsm, respectively, demonstrating that monostatic ISAC sensing enables environment-aware channel reconstruction with physical interpretability.

\section{Acknowledgements}
This work is supported by the MultiX and iSEE-6G projects under the EU’s Horizon Europe research and innovation programme (Grant no. 101192521 and 101139291). Z. Cui is supported by the Research Foundation – Flanders (FWO), Senior Postdoctoral Fellowship 12AFN26N.

\end{document}